\documentclass[
aps,%
12pt,%
final,%
notitlepage,%
oneside,%
onecolumn,%
nobibnotes,%
nofootinbib,%
superscriptaddress,%
centertags]%
{revtex4-1}
\usepackage[dvips]{graphicx}
\usepackage{amsmath,amssymb}
\begin{document}

\title{The Skyrme-TQRPA calculations of electron capture on hot nuclei in pre-supernova environment}

\author{\firstname{Alan~A.}~\surname{Dzhioev}}
\email{dzhioev@theor.jinr.ru} \affiliation{\rm Bogoliubov Laboratory
of Theoretical Physics, JINR, 141980 Dubna, Russia}

\author{\firstname{A.~I.}~\surname{Vdovin}}
\email{vdovin@theor.jinr.ru} \affiliation{\rm Bogoliubov Laboratory of Theoretical Physics, JINR,
141980 Dubna, Russia}

\author{\firstname{Ch.}~\surname{Stoyanov}}
\email{stoyanov@inrne.bas.bg} \affiliation{\rm Institute for Nuclear Research and Nuclear Energy,
boul. Tzarigradsko Shaussee 73, 1784 Sofia, Bulgaria}

\begin{abstract}
We combine the thermal QRPA approach with the Skyrme energy density
functional theory (Skyrme-TQRPA) for modelling the process of
electron capture on nuclei in supernova environment. For a sample
nucleus, $^{56}$Fe, the Skyrme-TQRPA approach is applied to analyze
thermal effects on the strength function of GT$_+$ transitions which
dominate electron capture at $E_e\le 30$~MeV. Several Skyrme
interactions are used in order to verify the sensitivity of the
obtained results to the Skyrme force parameters. Finite-temperature
cross sections are calculated and the results are compared with
those of the other model calculations.
\end{abstract}


\maketitle

\section{Introduction}

Now it is well established that weak-interaction processes with
nuclei play an important role in the dynamics of the collapsing core
of a massive star that leads to a supernova explosion
\cite{Langanke_RevModPhys75,Janka_PhysRep442}. During the
pre-collapse phase, the competition between electron capture (EC)
and $\beta$-decay determines the electron-to-baryon ratio ($Y_e$)
in the pre-supernova star  and hence its  Chandrasekhar mass
proportional to~$Y^2_e$. When the collapse proceeds, electron
capture reduces the number of electrons available for pressure
support, while $\beta$-decay acts in opposite direction. Until the
core reaches densities of $\rho\sim10^{11}~\mathrm{g~cm^{-3}}$,
neutrinos produced in this reaction leave the star freely, carrying
away energy and helping to maintain a low entropy. As a result,
nucleons resides primarily in nuclei. However, with increasing
densities neutrino interactions with matter become important and
influence the energy transfer from the core to the outer layers. So,
the supernova simulations should include all potentially important
weak-interaction processes and reliable estimates  of these rates
and cross sections would contribute to a better understanding of the
explosion mechanism.

In the present work, we focus our attention on electron capture. In
the stellar environment electron energies are typically less than
$30$~MeV and  at early stage of collapse EC is dominated by
Gamow-Teller (GT$_+$) transitions in iron-group nuclei ($A=45-65$).
Therefore the Gamow-Teller strength functions in iron-group nuclei
are of special importance. The task is complicated by the fact that
under extreme  conditions that hold in the supernova environment,
GT$_+$ transitions from thermally-populated excited states of the
parent nucleus may contribute significantly to EC. Unfortunately, to
obtain information about GT$_+$ transitions from excited states in
the terrestrial laboratory is not possible. Therefore, to describe
EC probabilities in supernovae we should rely on theoretical (model)
calculations.

Presently, the most reliable EC calculations for iron-group nuclei
are performed by using large-scale shell-model (LSSM)
diagonalization approach~\cite{Caurier_NPA653,Langanke_NPA481}. For
iron-group nuclei, present state-of-the-art shell model calculations
provides a detailed GT strength distribution for the nuclear ground
and excited states. However, for typical supernova temperatures
$T\approx 1$~MeV too many states can be thermally populated and this
makes state-by-state  evolution of the individual GT$_+$ strength
distributions computationally unfeasible. To overcome this problem
the Brink hypothesis is applied, i.e., it is assumed that GT$_+$
strength distributions on nuclear excited states are the same as for
the nuclear ground state. Thermal effects are treated by the
so-called back-resonance contribution (see~\cite{Langanke_NPA481}
for more details). However, the validity of Brink hypothesis for the
GT$_+$ strength function is not obvious and even more the
shell-model Monte-Carlo studies performed at finite
temperatures~\cite{Radha_PRC56} and the recent shell-model
calculations~\cite{Misch_PRC90} showed that the hypothesis is
failed.

To predict  EC rates and cross sections for hot nuclei, a so-called
thermal quasiparticle random-phase approximation (TQRPA) was
proposed recently in the framework of a statistical approach to the
nuclear many-body problem at finite temperature~\cite{Dzhioev_PAN72,
Dzhioev_PRC81}. In this approach, rather than computing individual
strength distributions for the nuclear ground and excited states,
one determines an "average" temperature dependent strength function.
In~\cite{Dzhioev_PAN72, Dzhioev_PRC81}, calculations were performed
for $^{54,56}$Fe and for neutron-rich germanium isotopes. The latter
can be considered as the average nucleus at later stages of
collapse~\cite{Cooperstein_NPA420}. It was found that the TQRPA does
not support the Brink hypothesis and leads to noticeable thermal
effects on the GT$_+$ strength function. As a result,  for the Ge
isotopes the low-energy cross sections are sensitive to temperature.
Later on, the method was also  applied to study neutrino-nucleus
reactions in supernova
environments~\cite{Dzhioev_PRC89,Dzhioev_PRC92} and similar thermal
effects on the low-energy cross section were found.

In~\cite{Dzhioev_PAN72, Dzhioev_PRC81}, the TQRPA calculations were
based on the Hamiltonian of the Quasiparticle-Phonon model
(QPM)~\cite{Soloviev1992} with a phenomenological Saxon-Woods
mean-field potential and schematic particle-hole interactions. The
parameters of  the QPM Hamiltonian were adjusted locally, i.e., to
properties of a nucleus under consideration. This feature strongly
reduces the predictive power of the theory.

In this paper, we extend our studies and perform self-consistent
calculations combining the TQRPA approach with the Skyrme energy
density functional theory. Use of the Skyrme forces makes more
reliable theoretical predictions of the nuclear properties far from
stability valley which play an important role in the process of
stellar collapse.

The present calculations are performed within the finite-rank
separable approximation, which expands the Skyrme residual
interaction into a sum of separable terms in a systematic
manner~\cite{Giai_PRevC57, Sever_PRC66, Sever_PTP128}. The
factorization considerably reduces the computational effort of the
TQRPA while maintaining high accuracy and even allows one to go
beyond the TQRPA. It should be mentioned that in
Refs.~\cite{Paar_PRC80,Fantina_PRC86} a finite-temperature RPA
(FTRPA) model based on Skyrme functionals has been already applied to
study EC in supernovae. Moreover, a similar approach, extended to
the relativistic framework (FTRRPA), has been employed in
Ref.~\cite{Niu_PRC83}. However, in the cited papers thermal effects
are treated not quite consistently. Below we discuss the subject in
more details and compare our results with those of Refs.
\cite{Paar_PRC80,Fantina_PRC86,Niu_PRC83}.

The paper is organized as follows. In Sec.~\ref{formalism}, we
briefly outline  the TQRPA formalism and the method of
separabilization of the
 Skyrme residual interaction. In Sec.~\ref{results}, the GT$_+$ thermal strength
functions and electron capture cross sections are presented for the
sample nuclei $^{56}$Fe. The results are compared  with those
obtained with the QPM Hamiltonian and within the FTRPA and FTRRPA
frameworks. In Sec.~\ref{conclusion}, we draw conclusions and give
an outlook for future studies. The derivation of the charge-exchange
TQRPA equations for the finite-rank separable Skyrme interaction is
given in  Appendix~\ref{appendix}.

\section{FORMALISM}\label{formalism}

\subsection{Thermal strength function}

During the core-collapse phase of a supernova explosion the
temperature  in the core is sufficiently high (a few $10^9$~K) to
establish an equilibrium of reactions mediated by the strong and
electromagnetic interactions~\cite{Janka_PhysRep442}. Neglecting
weak-interaction processes, one can consider nuclei as open quantum
systems in thermal equilibrium with the heat and particle reservoir
and, hence, they can be described as a thermal grand canonical
ensemble with temperature $T$ and proton and neutron chemical
potentials $\lambda_p$ and $\lambda_n$, respectively. Following
Refs.~\cite{Cooperstein_NPA420,Wambach1986}, to study EC on a hot
nucleus we  introduce a thermal strength function as a grand
canonical average of transition matrix elements of the GT$_+$
operator between states $i$ and $f$ in the parent and daughter
nuclei
\begin{equation}\label{tsf}
  S_{\mathrm{GT}_+}(E,T)=\sum_{Z,N}\sum_{i,f}S_{if}(\mathrm{GT}_+)\delta(E-E_{if}) P(i,A^Z_N).
\end{equation}
Here, $S_{if}(\mathrm{GT}_+) = |\langle f,A^{Z-1}_{N+1}|\sigma
t_{+}|i,A^Z_N\rangle|^2$  and $E_{if}=E_f-E_i+Q$ are,
respectively, the transition strength and the transition energy,
while $P(i, A^Z_N)$ determines the probability to  find the
initial state $i$ in the grand canonical ensemble. The $Q$ value is
the difference between the masses of the daughter and parent nuclei,
$Q=M_d-M_p$, and it determines the reaction threshold  at $T=0$.
Transition energy corresponds to the energy difference between the
incoming electron and the outgoing neutrino,  $E_{if} = E_e-E_\nu$.
At finite temperature, $E_{if}$ can take negative values due to
transitions from higher energy thermally excited states to lower
energy states.

For the EC cross section one has
\begin{align}\label{CrSect}
  \sigma(E_e,T) &= \frac{(G_F g_A)^2}{2\pi} F(Z,E_e)\int(E-E_e)^2 S_{\mathrm{GT}_+}(E,T)dE
  \notag\\
   &=\sigma_\mathrm{en}(E_e,T) + \sigma_\mathrm{ex}(E_e,T).
\end{align}
Here, $G_F$ is the weak interaction coupling constant, $g_A$ is the
axial coupling  constant, and $F(Z,E_e)$ is the Fermi function that
accounts for the Coulomb distortion of the electron wave function
near the nucleus (see, e.g., Ref.~\cite{Langanke_NPA481}). In
Eq.~\eqref{CrSect}, for further consideration, the total cross
section is split into two parts: $\sigma_\mathrm{en}(E_e,T)$
describes the endoergic process which requires an energy input and
includes only upward transitions ($E_{if}>0$), while
$\sigma_\mathrm{ex}(E_e,T)$ accounts for downward transitions
($E_{if}<0$) from thermally excited states and corresponds to the
exoergic process when EC releases energy. At $T\ne 0$, the latter
process is possible for arbitrary small incident electron energies,
i.e., there is no reaction threshold at finite temperatures.

To compute the thermal strength function we apply the thermal
quasiparticle  random-phase approximation (TQRPA) which is based on
the thermo-field dynamics (TFD) formalism. The concept of TFD is
expounded in~\cite{Umezawa1982,Takahashi_IJMPB10,Ojima_AnPhys137},
and here
 we just briefly outline the key points relevant for the present discussion.
In TFD, a hot nucleus is described by the state vector in the
doubled Hilbert  space  which is  a direct product of the original
space and its isomorphic tilde space. Such doubling of the system
degrees of freedom allows us to consider excitation and
de-excitation processes at finite temperature. The correspondence
between operators $A$ acting in the original Hilbert space and their
tilde-partners $\widetilde A$ is given by the tilde-conjugation
rules~\cite{Umezawa1982,Takahashi_IJMPB10,Ojima_AnPhys137}. The
important point is that the time evolution in the doubled Hilbert
space is generated by the thermal Hamiltonian
\begin{equation}\label{thermalH}
  \mathcal{H}=H-\widetilde H,
\end{equation}
where $\widetilde H= H(\widetilde a^\dag,\widetilde a)$ is the
tilde-partner  of the original nuclear Hamiltonian $H(a^\dag,a)$.
The zero-energy eigenstate $|0(T)\rangle$ of the thermal Hamiltonian
$\mathcal{H}$, which satisfies the thermal state condition
\begin{equation}\label{TSC}
A|0(T)\rangle = \sigma_A\,{\rm e}^{{\mathcal H}/2T} {\widetilde
A}^\dag|0(T)\rangle,
\end{equation}
is called the thermal vacuum and it describes the equilibrium state
of a hot nucleus.  Nonequilibrium states caused by an external
perturbation correspond to non-zero energy eigenstates of the
thermal Hamiltonian. By construction, the thermal Hamiltonian has
both positive- and negative-energy (tilde) eigenstates,
$\mathcal{H}|n\rangle=E_n|n\rangle$ and $\mathcal{H}|\widetilde
n\rangle=-E_n|\widetilde n\rangle$.

Given the eigenstates of the thermal Hamiltonian, the thermal
strength function for  any transition operator $\cal T$ can be
written as
\begin{equation}\label{str_func_TFD}
  S_\mathcal{T}(E,T)=\sum_n\bigl\{ S_n(\mathcal{T})\delta(E-E_n)+\widetilde S_n(\mathcal{T})\delta(E+E_n)\bigr\},
\end{equation}
where $S_n(\mathcal{T})$ and $\widetilde S_n(\mathcal{T})$ are the
transition  strengths from the thermal vacuum
\begin{align}
  S_n(\mathcal{T}) = |\langle n|\mathcal{T}|0(T)\rangle|^2,
  \notag\\
  \widetilde S_n(\mathcal{T}) = |\langle\widetilde n|\mathcal{T}|0(T)\rangle|^2.
\end{align}
Obviously, in most practical cases one cannot diagonalize
$\mathcal{H}$ exactly.  In the present study, to obtain the thermal
GT$_+$ strength function, we apply the TQRPA. In this method,
nonequilibrium states of a hot nucleus are treated as phonon-like
excitations on the thermal vacuum. Thus, the problem is reduced to
the diagonalization of the thermal Hamiltonian in terms of phonon
operators such that the respective phonon vacuum obeys the
thermal-state condition~\eqref{TSC}.  Below we briefly outline the
method, while the details can be found
in~\cite{Dzhioev_IJMPE18,Dzhioev_PRC81} and in
Appendix~\ref{appendix}.

\subsection{Proton-neutron TQRPA with finite-rank separable approximation for the Skyrme interaction}

To obtain the thermal GT$_+$ strength function within the TQRPA we
suppose that  the nuclear proton and neutron Hartree-Fock states are
already produced using the Skyrme energy density functional. In
particular, it means that we ignore the influence of temperature on
the nuclear mean field. Following~\cite{Bortignon_1998}, this
stability of the mean field with respect to temperature is expected
for $T$ considerably smaller than the energy difference between
major shells ($\hbar\omega_0 = 41A^{-1/3})$. This requirement is
well satisfied in nuclei with $A < 100$ for the maximum temperatures
reached during the collapse ($T\sim 5$~MeV). Thus, the model
Hamiltonian has the form
\begin{equation}
  H = H_\mathrm{mf}+H_\mathrm{pair}+H_\mathrm{ph}
\end{equation}
and it contains a spherical Skyrme-HF mean field for nucleons, the
pairing interaction  and the residual particle-hole interaction.
Since we are working in the grand-canonical ensemble, the chemical
potentials $\lambda_n$ and $\lambda_p$ are  included into
$H_\mathrm{mf}$. The particle-hole interaction $H_\mathrm{ph}$ is
defined in terms of second derivatives of the Skyrme energy density
functional with respect to the one-body
density~\cite{Bertsch_PRepC18} and  can be written in terms of the
Landau-Migdal theory of Fermi systems. Keeping only $l=0$ terms in
$H_\mathrm{ph}$, the isovector part of the residual interaction
which is responsible for charge-exchange excitations  reads
\begin{equation}\label{S_Landau}
  H_\mathrm{ph}=N_0^{-1}[F^\prime_0+ G_0^\prime\sigma_1\cdot\sigma_2]\tau_1\cdot\tau_2\delta(\mathbf{r}_1-\mathbf{r}_2),
\end{equation}
where $\sigma$ and $\tau$ are the nucleon spin and isospin
operators,  and $N_0=2k_Fm^*/\pi^2\hbar^2$ with $k_F$ and $m^*$
denoting the Fermi momentum and nucleon effective mass,
respectively. The expressions for the Landau parameters
$F^\prime_0$, $G^\prime_0$ in terms of the Skyrme force parameters
can be found in Ref.~\cite{Giai_PLB379}.  Here we just mention that
due to the density dependence of the Skyrme interaction, the Landau
parameters are functions of the coordinate~$\mathbf{r}$.

Following the method presented in~\cite{Giai_PRevC57,Sever_PRC66},
we apply an  $N$-point integration Gauss formula and reduce the part
of $H_\mathrm{ph}$ responsible for charge-exchange excitations  to a
finite-rank separable form
\begin{equation}
   H^\mathrm{ch}_\mathrm{ph} = -2\sum_{k=1}^{N} \sum_{JM}\kappa^{(k)}_F M^{(k)\dag}_{JM} M^{(k)}_{JM} -2\sum_{k=1}^{N} \sum_{LJM}\kappa^{(k)}_G S^{(k)\dag}_{LJM} S^{(k)}_{LJM}.
   \end{equation}
The isovector interaction strengths, $\kappa^{(k)}_F$ and
$\kappa^{(k)}_G$, are expressed via the Landau
parameters~\cite{Giai_PRevC57,Sever_PRC66}. The multipole and
spin-multipole operators entering $H_\mathrm{ph}$ are given
by\footnote{In~\eqref{MS} and hereinafter, $[~]^J_M$ denotes the
coupling of two single-particle angular momenta $j_p$, $j_n$ to the
angular momentum $J$. The bar over index $j$ implies time
inversion.}
\begin{align}\label{MS}
 \hat M^{(k)\dag}_{JM} &=  \hat J^{-1}\sum_{j_pj_n} f^{(Jk)}_{j_pj_n}[a^\dag_{j_p}a_{\overline{\jmath_n}}]^J_M,
  \notag\\
 \hat S^{(k)\dag}_{JM} &=  \hat J^{-1}\sum_{j_pj_n} g^{(LJk)}_{j_pj_n}[a^\dag_{j_p}a_{\overline{\jmath_n}}]^J_M,
\end{align}
where $\hat J=\sqrt{2J+1}$ and $f^{(Jk)}_{j_pj_n}$, $g^{(LJk)}_{j_pj_n}$ denote the reduced single-particle matrix elements
\begin{align}
f^{(Jk)}_{j_pj_n}&=u_{j_p}(r_k)u_{j_n}(r_k)\langle j_p\|i^J Y_J\|j_n\rangle
\notag\\
g^{(LJk)}_{j_pj_n}&=u_{j_p}(r_k)u_{j_n}(r_k)\langle j_p\|i^L[Y_L\times\sigma]^M_J\|j_n\rangle.
\end{align}
The radial wave functions $u_j(r_k)$ are related to the Hartree-Fock
single-particle  wave functions~\cite{Giai_PRevC57,Sever_PRC66},
while $r_k$ are abscissas used in the $N$-point integration Gauss
formula.

Following the TFD, to study charge-exchange excitations in a hot
nucleus we should  double the original nuclear degrees of freedom by
introducing the tilde creation and annihilation operators
$\widetilde a^\dag_{jm},~\widetilde a_{jm}$ and then diagonalize the
respective thermal Hamiltonian~\eqref{thermalH}.
 Within the TQRPA, the thermal Hamiltonian is diagonalized in two steps. First, we
 introduce thermal quasiparticles that diagonalize the mean field and pairing parts of $\mathcal{H}$
\begin{equation}
  \mathcal{H}_\mathrm{mf}+\mathcal{H}_\mathrm{pair} \simeq \sum_{\tau=n,p}{\sum_{jm}}^\tau \varepsilon_{jm}(T)(\beta^\dag_{jm}\beta_{jm}-
  \widetilde{\beta}^\dag_{jm}\widetilde{\beta}_{jm}).
\end{equation}
Thermal quasiparticles account pairing correlations at finite
temperature and their  energy and structure are found from the
finite temperature BCS equations (see. Ref.~\cite{Dzhioev_PRC81} for
more details). In accordance with the BCS
theory~\cite{Goodman_NPA352, Civitarese_NPA404}, the numerical
solution of these equations yields vanishing of pairing correlations
above a certain critical temperature~$T_\mathrm{cr}$.

The next step is to account for the residual particle-hole interaction
and diagonalize  the thermal Hamiltonian in terms of thermal phonon
creation and annihilation operators
\begin{equation}
{\cal H}\simeq\sum_{JM i}\omega_{J i}(T)
   (Q^\dag_{JM i}Q^{\phantom{\dag}}_{JM i}
   -\widetilde Q^\dag_{JM i}\widetilde Q^{\phantom{\dag}}_{JM i}).
\end{equation}
The energies and structure of thermal charge-exchange phonons are
obtained by  the solution of TQRPA equations. The explicit form of
the proton-neutron TQRPA equations for the finite-rank separable
Skyrme forces is given in the Appendix.

Charge-exchange GT$_+$ transitions from the thermal vacuum result in
$J^\pi=1^+$  thermal phonon states. Once the energies and structure
of $1^+$ thermal phonons are determined, one can evaluate the
thermal strength function~\eqref{str_func_TFD} for GT$_+$
transitions. The  transition strengths are given by the following
reduced matrix elements\footnote{Explicit expressions for transition
strengths  can be found in~\cite{Dzhioev_PAN72, Dzhioev_PRC81}}
\begin{align}\label{trans_ampl}
S_i(\mathrm{GT}_+)&=\bigl|\langle Q_{1^+i}\|\sigma t_+\|0(T)\rangle\bigr|^2,
  \notag\\
\widetilde S_i(\mathrm{GT}_+)&=\bigl|\langle \widetilde Q_{1^+i}\|\sigma t_+\|0(T)\rangle\bigr|^2,
\end{align}
while the respective transition energies are
\begin{align}\label{TrEn}
  &E_i  = \omega_{Ji}+\delta_{np},
  \notag\\
  &\widetilde  E_i = -\omega_{Ji}+\delta_{np},
\end{align}
where $\delta_{np}=\lambda_n-\lambda_p + \Delta M_{np}$, and $\Delta
M_{np}=1.29$~MeV  is the neutron-proton mass difference (the
contribution $\delta_{np}$ arises because in charge-exchange
reactions the initial and final nucleons are attached to different
nucleon systems). Thus, within the TQRPA we have both positive- and
negative-energy transitions to thermal phonon states. The latter
contribute to exoergic EC when $E_\nu>E_e$. It should be emphasized
that in the zero-temperature limit, transition strengths to
tilde-phonon states vanish and the TQRPA method reduces into the
standard QRPA. In particular, at $T=0$ the transition energies
$E_i=\omega_{Ji}+\lambda_n-\lambda_p + \Delta M_{np}$
 correspond to the excitation energies with respect to the parent nucleus ground state.

In concluding this section we would like to point out that
the  thermal  strength function for  GT$_-$ transitions can
be obtained by the same method.
 In~\cite{Dzhioev_PAN72}, it was shown that within the TQRPA the total GT$_-$ and GT$_+$
 strengths fulfill the Ikeda sum rule
 \begin{equation}
   S_- - S_+ = 3(N-Z),
 \end{equation}
where $S_\mp = \int S_{\mathrm{GT}_\mp} dE$. Moreover, the GT$_-$ and GT$_+$
strength functions are related  by the detailed balance
\begin{equation}\label{db}
  S_{\mathrm{GT}_-}(-E,T) = S_{\mathrm{GT}_+}(E,T)\exp\left\{-\frac{E-\delta_{np}}{T}\right\}.
\end{equation}
Thus, for each $n\to p$ ($p\to n$) GT transition with energy $E>0$
there is an  inverse $p\to n$ ($n\to p$) transition with energy $-E$
and the respective transition strengths are connected by~\eqref{db}.
In~\cite{Dzhioev_PRC92}, we have shown in a model-independent way
that the  relation~\eqref{db} is valid in the grand-canonical
ensemble for any transition operators $\cal{T}_-$ and  $\cal{T}_+$,
which differ only by the isospin operator.

\section{Results}\label{results}

In this section, we employ the theoretical framework described above
to compute  EC cross sections on  $^{56}\mathrm{Fe}$ at finite $T$.
Experimental data available for this nucleus allow to test our
calculations at zero temperature. Besides, EC calculations for
$^{56}\mathrm{Fe}$ in the supernova environment have been performed
within various theoretical approaches
\cite{Paar_PRC80,Fantina_PRC86,Niu_PRC83} and these results can be
compared with those of the TQRPA. To reveal the sensitivity of the
results to the Skyrme interaction parametrization, we perform the
calculations for a representative set of Skyrme forces:
Sly4~\cite{Chabanat_NPA635}, SGII~\cite{Giai_PLB379}, and
SkM*~\cite{Bartel_NPA79}. To distinguish the present results from
those obtained with the QPM
Hamiltonian~\cite{Dzhioev_PRC81,Dzhioev_PRC92}, we will refer to
them as the Skyrme-TQRPA and the QPM-TQRPA, respectively.

A short comment should be made concerning the choice of the pairing
interaction.  Within the BCS approach the phase transition in nuclei
from the superfluid to normal state occurs at critical temperature
$T_\mathrm{cr}\approx0.5\Delta$, where $\Delta$ is the ground state
pairing gap~\cite{Goodman_NPA352,Civitarese_NPA404}. Therefore, the
inclusion of particle-particle residual interactions does not affect
the strength function for temperatures $T> T_\mathrm{cr}$. However,
to  compute the ground state GT$_+$ distributions and compare them
with the experimental and shell-model ones, pairing correlations are
taken into account at zero temperature. As
in~\cite{Dzhioev_PRC81,Dzhioev_PRC92} we employ presently a BCS
Hamiltonian with a constant pairing strength. The neutron and proton
pairing strength parameters are fixed to reproduce the odd-even mass
difference. At $T=0$ the resulting proton and neutron energy gaps
for $^{56}$Fe are $\Delta_{p}=1.57$~MeV and $\Delta_{n}=1.36$~MeV,
respectively. Thus, the critical temperature  when the pairing phase
transition occurs is $T_\mathrm{cr}\approx 0.8$~MeV.

\subsection{GT$_+$ strength function at zero and finite temperatures}

\begin{figure*}[h!]
\includegraphics[width=8cm]{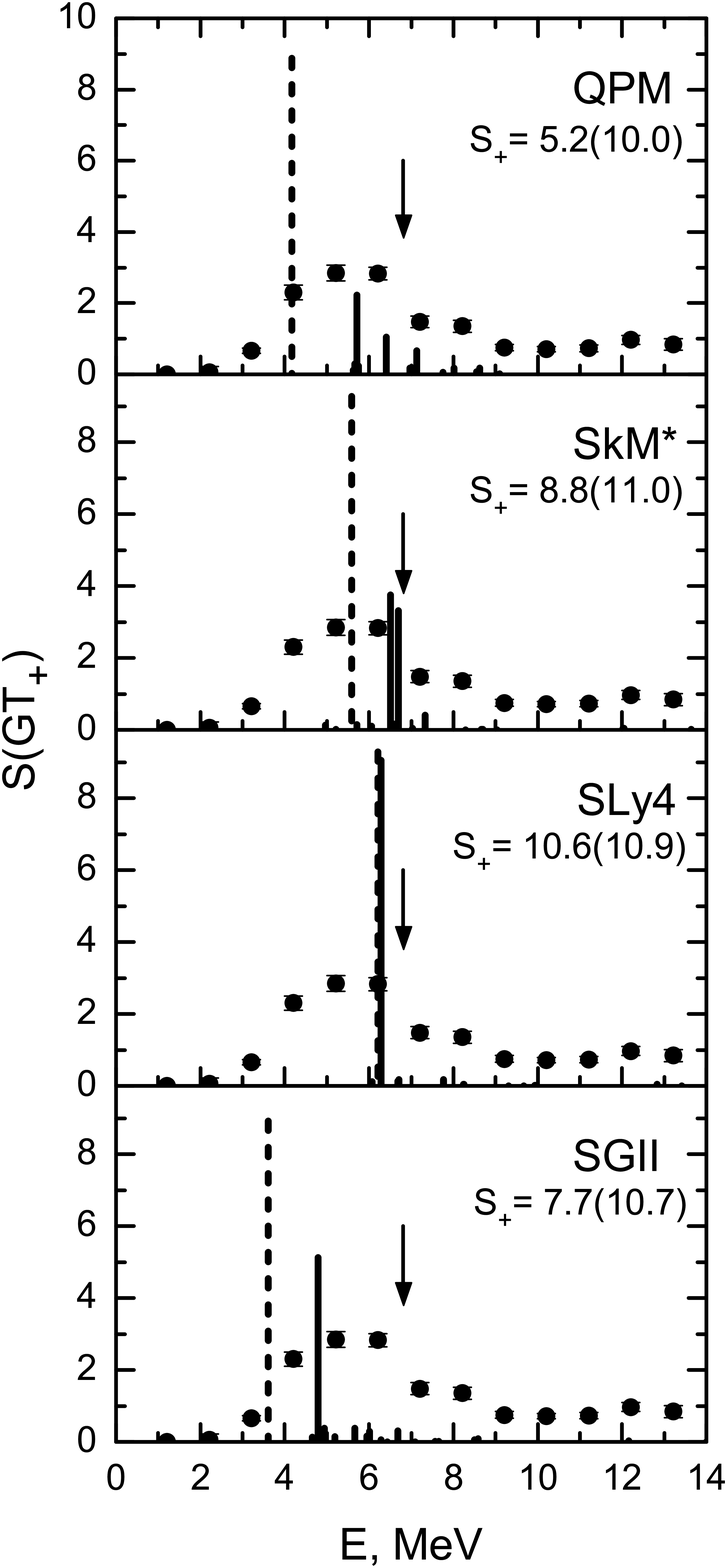}
 \caption{ Ground state GT$_+$ strength distributions in
$^{56}$Fe calculated with the SGII, SLy4 and SkM* forces. The
excitation energies are related to the parent ground state. For
comparison, the GT$_+$ strength calculated  with the QPM Hamiltonian
is also shown~\cite{Dzhioev_PRC81}. The solid peaks denote the QRPA
results, and the  dashed peaks represent the unperturbed BCS
distributions calculated neglecting the residual p-h interaction.
The total GT$_+$ strength is denoted by $S_+$ and the unperturbed
values of $S_+$ are given in parentheses. Experimental
data~\cite{ElKateb_PRC49} are displayed by points and for clearer
presentation they are multiplied by a factor of 5. The GT$_+$
centroid energy from the LSSM calculation~\cite{Caurier_NPA653}  is
indicated by an arrow. }\label{figure1}
\end{figure*}

In this subsection, we discuss temperature evolution of the  GT$_+$
strength function in $^{56}$Fe. To begin with, let us first consider
the results of QRPA calculations at zero temperature. In
Fig.~\ref{figure1}, we show the ground state  GT$_+$ strength
distribution,  whose measurement is feasible from
$(n,p)$~\cite{ElKateb_PRC49} reactions on the $^{56}$Fe target.
Notice that all distributions are plotted as functions of the
excitation energy with respect  to the parent nucleus ground state.
The experimental data from Ref.~\cite{ElKateb_PRC49} are indicated
by points and for convenience of comparison with the QRPA results
they are multiplied by a factor of 5. The GT$_+$ centroid energy,
6.81~MeV, predicted by the LSSM calculations~\cite{Caurier_NPA653}
is shown by an arrow (to obtain this number we have added the mass
splitting between daughter and parent nucleus,
$M(^{56}\mathrm{Mn})-M(^{56}\mathrm{Fe})=4.21$~MeV, to the number in
Table~1 of Ref.~\cite{Caurier_NPA653}).

Looking at Fig.~\ref{figure1} one can see that the structure of the
GT$_+$ strength  distributions is qualitatively similar for all the
Skyrme forces. Namely, our Skyrme-QRPA calculations produce strength
distributions mainly concentrated in a single resonance peak. The
peak is dominated by the single-particle transition $\pi1f_{7/2}\to
\nu1f_{5/2}$. Although the resonance is displaced in energy for the
different Skyrme interactions, the deviation of the main theoretical
peak from the maximum of experimental strength distribution lies
within 1\,MeV. When comparing the resonance energy with the LSSM
results, we notice that the QRPA calculations with SLy4 and SkM*
fairly well reproduce the GT$_+$ energy centroid predicted by the
shell-model calculations. Of course, the simple QRPA calculations
cannot reproduce the fragmentation of the strength, that is, the
spreading width. In this respect the LSSM
calculations~\cite{Caurier_NPA653} are clearly advantageous.

Figure~\ref{figure1} also shows the unperturbed GT$_+$ strength
distributions obtained  within the BCS approach, i.e., neglecting
the particle-hole residual interaction $H_\mathrm{ph}$. As evident
from the figure, the particle-hole interaction pushes the GT$_+$
strength to higher energies and the energy shift is the greatest for
the QPM-QRPA calculations. Moreover, due to particle-hole
correlations the GT$_+$ strength distribution calculated within
QPM-QRPA is more fragmented. At the same time, the BCS and QRPA
calculations with the SLy4 force produce practically the same
strength distributions. It means that for the SLy4 Skyrme force the
p-h residual interaction in spin-isospin channel is very weak. Not
only the resonance energy, but also the total GT$_+$ strength $S_+$
is affected by the residual interaction. Within the BCS, all
calculations predict rather close values of $S_+\approx 10\div11$.
The particle-hole correlations reduce the total GT$_+$ strength and
this effect is most significant for the QPM  based calculations.
However, despite the reduction, the  QRPA values of $S_+$ noticeably
overestimate the experimental ones ($S_+=2.9\pm0.3$
\cite{ElKateb_PRC49}). The experimentally observed quenching of the
total GT strength is usually reproduced by reducing the axial
coupling constant from its free-nucleon value $g_A=-1.26$ to some
effective value $g^*_A$. In what follows we will use $g^*_A=-0.93$,
that corresponds to renormalization of the GT matrix elements by a
quenching factor $0.74$. The same quenching factor was used in the
shell-model calculations~\cite{Caurier_NPA653}.

\begin{figure*}[h!]
\includegraphics[width = 8cm]{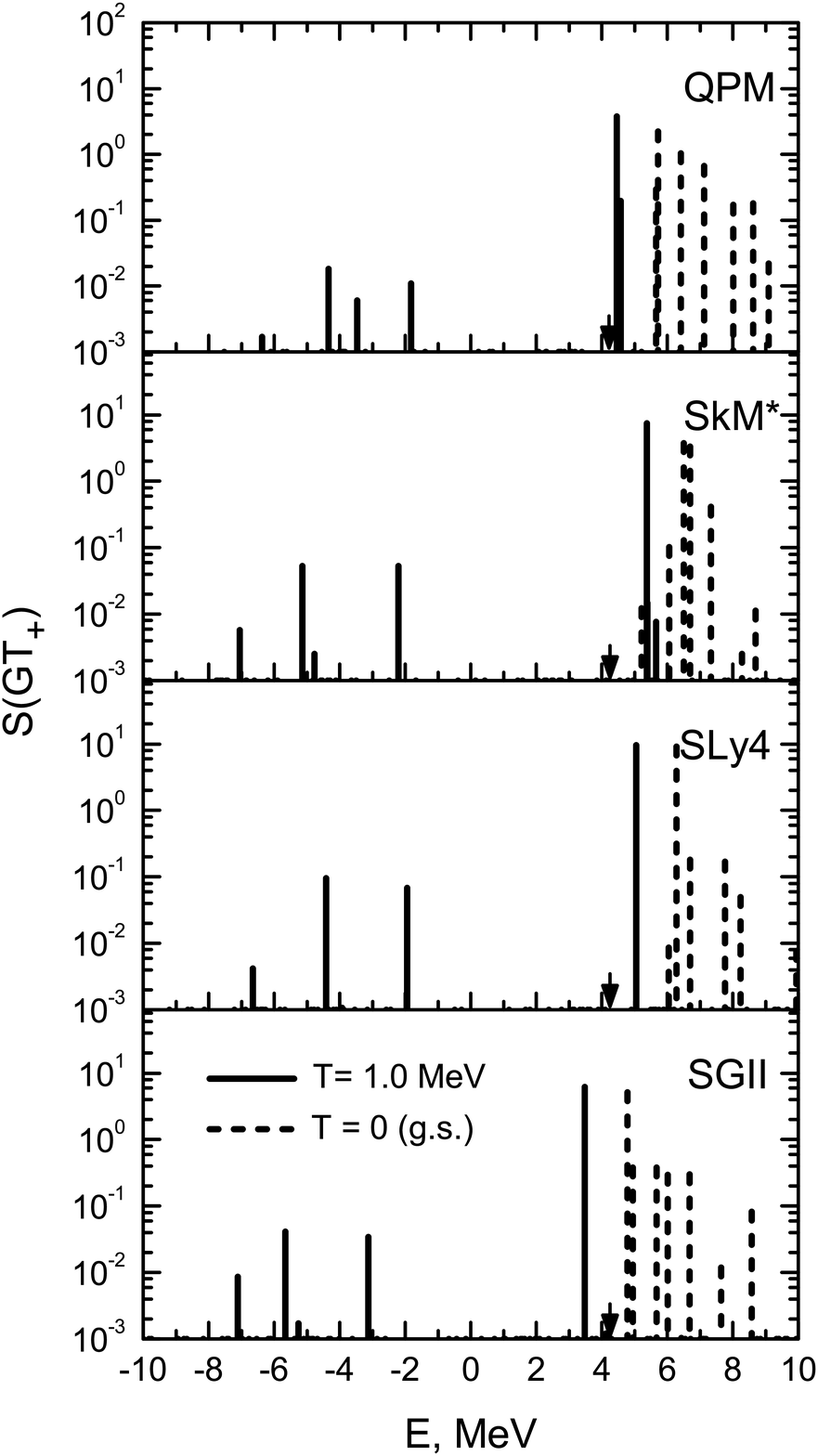}
 \caption{GT$_+$ strength functions for  $^{56}$Fe
calculated at $T=0$ (dashed peaks) and $T=1.0$~MeV (solid peaks).
The arrows indicate the ground-state reaction threshold for the
electron capture ($Q=4.21$~MeV) }\label{figure2}
\end{figure*}

Let us now compare the results of
Ref.~\cite{Dzhioev_PRC81,Dzhioev_PRC92} where thermal  effects on
the GT$_+$ strength function where studied within the QPM-TQRPA
approach with the present self-consistent scheme based on the Skyrme
energy density functional theory. The GT$_+$ thermal strength
function in $^{56}$Fe is shown in Fig.~\ref{figure2} at $T=1$~MeV.
To make the thermal effects clearly defined the ground state ($T=0$)
strength functions are also shown. Note, that the strength functions
are displayed in a logarithmic scale.

In Fig.\,2, one can easily see that the Brink hypothesis is not
valid for hot nuclei and the GT$_+$ strength function evolves with
temperature. Effective interaction affects this thermal effect
quantitatively but not qualitatively. For the upward ($E>0$)
strength, the main effect is a temperature-induced shift of the
GT$_+$ resonance towards lower energies. This decrease is mainly
attributed to the vanishing of pairing correlations, since at
temperatures above the critical one no extra energy  is needed to
break a proton Cooper pair when performing GT$_+$ transitions.   Our
QPM-TQRPA and Skyrme-TQRPA calculations show that when the
temperature is increased up to 1~MeV, the  GT$_+$ resonance is
lowered by about~$1.5$~MeV. In particular, calculations with the
SGII force demonstrate that due to pairing collapse the GT$_+$
resonance shifts below the ground-state reaction threshold
$Q=M(^{56}\mathrm{Mn})-M(^{56}\mathrm{Fe})$. However, not only
vanishing of pairing correlations causes the resonance downward
shift. It was shown in~\cite{Dzhioev_PRC92}, that owing to the
thermal blocking of the residual interaction, a further increase in
temperature could decrease the  GT$_+$ resonance as well. As
mentioned in the Introduction, the observed temperature-induced
downward shift of the GT$_+$ resonance is not present in LSSM
calculations, since they are partially based on the Brink
hypothesis. In contrast, the finite temperature relativistic QRPA
calculations~\cite{Niu_PRC83} and shell-model Monte-Carlo
calculations~\cite{Radha_PRC56} show similar features for the
changes of the GT$_+$ resonance energy.

At finite temperature, GT$_+$ transitions which are Pauli blocked at
$T=0$   due to closed neutron subshell become unblocked due to
thermal smearing of the nuclear Fermi surface. Similarly, protons
that are thermally excited to higher orbitals can undergo GT$_+$
transitions. In $^{56}$Fe, such thermally unblocked transitions lead
to appearance of the downward ($E<0$) component in the GT$_+$
strength function. It is interesting to note that both the QPM-TQRPA
and Skyrme-TQRPA calculations predict roughly the same energy region
where the thermally unblocked GT$_+$ strength appears at $T=1$~MeV.
The single-particle transitions which mainly contribute to this
strength are $\pi 2p_{3/2}\to \nu 2p_{3/2,1/2}$ particle-particle
and $\pi 1f_{7/2}\to\nu 1f_{7/2}$ hole-hole transitions. Here
particle (hole) denotes a state above (below) the Fermi level.

It should be emphasized that the appearance of downward transitions
in the TQRPA  thermal strength function stems from the doubling of
the system degrees of freedom within the TFD. For $^{56}$Fe, this
downward strength corresponds to transitions to tilde-phonon states,
i.e., to negative-energy solutions of the TQRPA equations. No such
negative-energy transitions appear within the approaches based  on
the  finite temperature RPA used in~\cite{Paar_PRC80, Niu_PRC83,
Fantina_PRC86}. Therefore, only upward GT$_+$ transitions were
considered in calculations of the EC rates on $^{56}$Fe within the
FTRPA and FTRRPA.

\subsection{Electron capture cross section}

\begin{figure*}[h!]
\includegraphics[width=17cm]{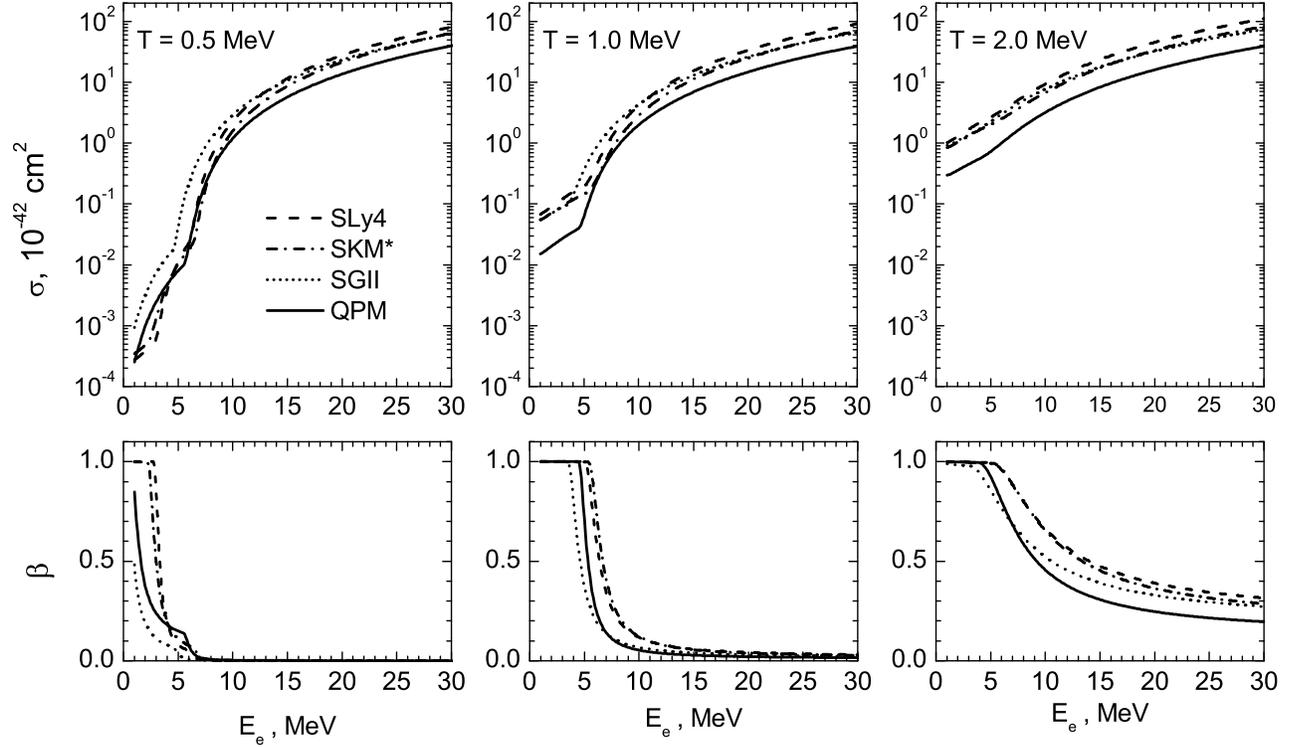}
\caption{ Top panels: Electron capture cross sections for
$^{56}$Fe for three different temperatures. The Skyrme-QRPA results
are compared with those  obtained by the QPM-TQRPA calculations.
Bottom panels: Temperature dependence of the ratio $\beta(E_e,T)$ of
the exoergic electron absorption to the reaction cross section. }\label{figure3}
\end{figure*}

In Fig.~\ref{figure3}, we display the calculated EC cross
sections~\eqref{CrSect} as functions of the incident electron energy
$E_e$. The cross sections are shown
 at three different temperatures, $T=0.5,~1.0$ and 2.0~MeV. Moreover, the Skyrme-TQRPA
results are presented together with those of the QPM-TQRPA
calculations. As seen from the plots, all models predict a universal
behavior of the cross section  versus electron energy and
temperature. In particular, there is no reaction threshold for EC at
finite temperature and the low-energy cross sections demonstrate a
significant thermal enhancement. It is clear that both these effects
are caused by  downward GT$_+$ transitions from thermally excited
states which contribution to the cross section increases with
temperature.

The bottom panels of Fig.~\ref{figure3} show the ratio of exoergic
EC to the reaction cross section
   \begin{equation}
     \beta(E_e,T) =\frac{ \sigma_\mathrm{ex}(E_e,T)}{\sigma(E_e,T)}.
   \end{equation}
As expected, the ratio $\beta \sim 1$   for low-energy electrons and
then gradually decreases with increasing electron energy. Moreover,
the higher the temperature the wider is the range of $E_e$ when
exoergic process dominates (i.e., $\beta>0.5$) EC. It should be
stressed that all variants of the Skyrme forces used here give
rather similar results. The spread in calculated cross sections is
less than an order of magnitude at low energies and temperatures and
it decreases with the increase of $T$ and $E_e$. The Skyrme-TQRPA
calculations systematically predict cross sections above the values
obtained within the QPM-TQRPA model. Evidently, the discrepancy
reflects the differences in the total GT$_+$ strength (see $S_+$
values in Fig.~\ref{figure1}).

\begin{figure*}[h!]
\includegraphics[width=17 cm]{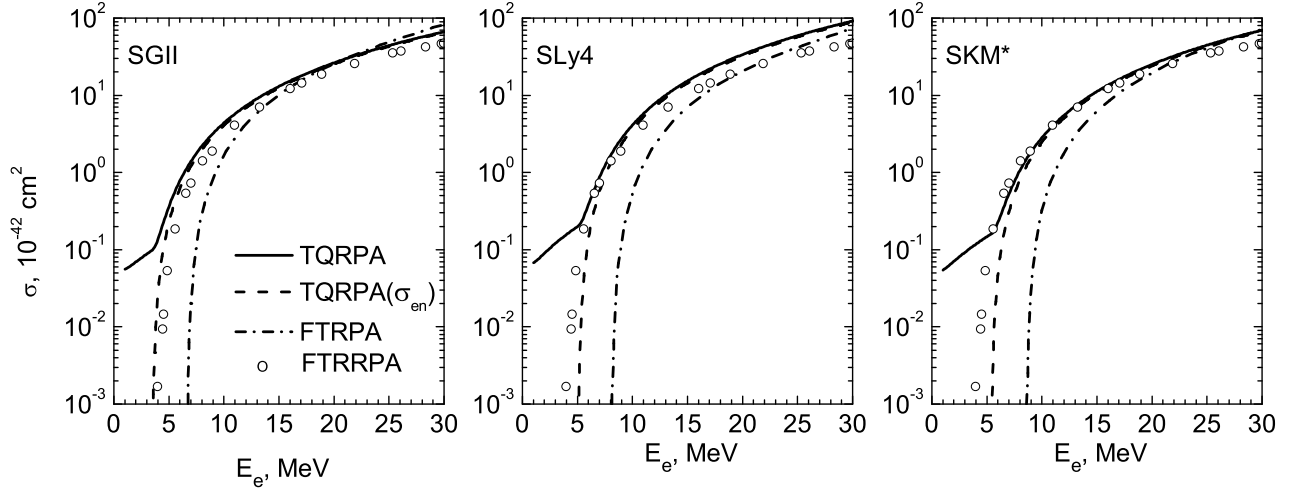}
 \caption{Electron capture cross sections for $^{56}$Fe at
$T=1.0$~MeV.  The Skyrme-TQRPA results are compared with the cross
sections calculated within the FTRPA~\cite{Fantina_PRC86} and the
FTRRPA~\cite{Niu_PRC83} framework }\label{figure4}
\end{figure*}

In Fig.~\ref{figure4}, the present results of the Skyrme-TQRPA
calculations at $T=1$~MeV  are compared with those obtained by the
FTRPA~\cite{Fantina_PRC86} and the FTRRPA~\cite{Niu_PRC83}
calculations. In each plot we compare the TQRPA and FTRPA cross
sections calculated with the same Skyrme force. One can notice in
the figure that the FTRPA and FTRRPA calculations predict the cross
section rapidly dropping to zero when the electron energy tends to
some threshold value.  As was pointed above, the FTRPA and FTRRPA
approaches do not include downward GT$_+$ transitions that
contribute to the exoergic EC. For this reason some minimal electron
energy is required to trigger the EC process. In contrast, within
the TQRPA, downward transitions dominate the low-energy cross
section at $T=1.0$~MeV and make possible EC for arbitrary small
incident electron energy.

In Fig.~\ref{figure4}, we also display the endoergic component of
the cross section calculated with the Skyrme-TQRPA. As seen, the
general behavior of $\sigma_\mathrm{en}(E_e,T)$ as a function of
$E_e$ is in agreement with the FTRPA and TQRPA calculations.
However, the TQRPA results are much closer to those computed within
the FTRRPA framework than to the FTRPA results obtained with the
same Skyrme forces. Namely, the FTRPA cross sections are shifted to
higher electron energies with respect to our
$\sigma_\mathrm{en}(E_e,T)$ and  the shift is practically the same
($\sim 3$~MeV) for all the Skyrme forces used. It seems that the
discrepancy reflects the difference in the GT$_+$ peak position and
the reason for this most likely lies in the different definition of
transition energies. To explain this, we recall that in the FTRPA it
is assumed\footnote{See Eq.~(4) in \protect
Ref.~\cite{Fantina_PRC86} and discussion therein.} that the RPA
energy corresponds to the excitation energy in the daughter nucleus.
In such a case, one can approximately write the transition energy to
the GT$_+$ resonance in $^{56}$Fe as
\begin{equation}
  E_\mathrm{FTRPA} = E(\nu 1f_{5/2}) - E(\pi 1f_{7/2}) + \Delta H_{ph} + Q,
\end{equation}
where $Q=M(^{56}\mathrm{Mn})-M(^{56}\mathrm{Fe})=4.21$~MeV and
$\Delta H_{ph}$  is the energy shift induced by the residual
interaction. In the TQRPA framework, the transition energy is
defined by Eq.~\eqref{TrEn} and it can be written as
\begin{align}
  E_\mathrm{TQRPA} &= \varepsilon(\nu 1f_{7/2}) + \varepsilon(\nu 1f_{7/2}) + \Delta H_{ph} + \delta_{np}
  \notag\\
  & = E(\nu 1f_{5/2}) - E(\pi 1f_{7/2}) + \Delta H_{ph}+\Delta M_{np}.
\end{align}
Here, we take into account that in the absence of pairing
correlations  single-particle and quasiparticle energies are
connected as $\varepsilon = \pm(E-\lambda)$, where the upper sign is
for particle states, and the lower sign is for hole states. Thus,
one has $ E_\mathrm{FTRPA} - E_\mathrm{TQRPA}=2.92~\mathrm{MeV}$
which is very close to the observed  energy shift between the FTRPA
cross section and our $\sigma_\mathrm{en}(E_e,T)$. It should be also
mentioned that under the FTRRPA  (see Eq.~(14) in
Ref.~\cite{Niu_PRC83}) the transition energy $E_e-E_\nu$ is
determined by the same manner as under the TQRPA (at
$T>T_\mathrm{cr}$). Therefore, it is not surprising that the FTTRPA
cross section is in a good agrement with our
$\sigma_\mathrm{en}(E_e,T)$.

\section{Summary and conclusions}\label{conclusion}

In the present work the electron capture cross sections on the hot
$^{56}$Fe nucleus were calculated in the supernova environment. The
thermal effects were treated within the thermal QRPA combined with
Skyrme energy density functional theory. The results were compared
with those obtained from the TQRPA calculations with the QPM
Hamiltonian as well as from the finite temperature RPA and relativistic
RPA approaches (see Ref.~\cite{Fantina_PRC86} and
Ref.\,\cite{Niu_PRC83}, respectively).

We perform a detailed analysis of thermal effects on the GT$_+$
transitions which dominate  the EC cross sections for
$E_e\le30$~MeV. It was found that the self-consistent TQRPA
calculations with the Skyrme forces predict the same thermal effects
on the GT$_+$ strength function as those found in our previous
studies based on the QPM Hamiltonian. In particular, increasing
temperature shifts the GT$_+$ resonance to lower energies and makes
possible negative-energy transitions. The values of the resonance
shift and the energies of thermally unblocked downward transitions
well agree for all Skyrme-TQRPA calculations.

We calculate the EC cross sections for different supernova
temperatures. The spread in the cross sections computed with the
different Skyrme forces is less than an order of magnitude. This
finding is the main result of the present study. Comparison with the
FTRPA and FTRRPA results [17,18] reveals the importance of downward
GT$_+$ transitions  for the low-energy EC cross section.

The application of the present self-consistent method is not
restricted by  iron-group nuclei and it can be applied to more
massive neutron-rich nuclei, which are beyond the present capability
of the LSSM calculations.   The fragmentation of the GT$_+$ strength
plays a significant role at low temperature and densities of the
supernova environment. Therefore, a further improvement of the model
is to go beyond TQRPA and take into account higher-order
correlations. For the separable residual interaction this can be
done by coupling the thermal phonon states with more complex (e.g.,
two-phonon) configurations. For charge-exchange excitations at zero
temperature, the phonon coupling was considered within the QPM
model~\cite{Kuzmin_JPG10}, and most recently with the
self-consistent Skyrme based calculations~\cite{Sever_RJPh58}.
Another possible improvement is the inclusion of the effect of
nuclear deformation. In~\cite{Sarriguren_PRC87}, EC calculations for
deformed nuclei were performed assuming that the process is
dominates by the ground-state contribution.

\section*{Acknowledgments}

We thank Dr.~A.~P.~Severyukhin  for
fruitful discussions.

\appendix
\section{} 
\label{appendix}

Within the TQRPA, charge-exchange thermal phonons are defined as a linear superposition of the proton-neutron thermal quasiparticle pair creation and annihilation operators
\begin{multline}\label{ch_phonon}
 Q^\dag_{JM i}=\sum_{j_p j_n}
  \Bigl(
  \psi^{J i}_{j_pj_n}[\beta^\dag_{j_p}\beta^\dag_{j_n}]^J_M+
  \widetilde\psi^{J i}_{j_pj_n}[\widetilde\beta^\dag_{\overline{\jmath_p}}
  \widetilde\beta^\dag_{\overline{\jmath_n}}]^J_M+
  i\eta^{J i}_{j_p j_n}[\beta^\dag_{j_p} \widetilde\beta^\dag_{\overline{\jmath_n}}]^J_M+
  i\widetilde\eta^{J i}_{j_pj_n}[\widetilde\beta^\dag_{\overline{\jmath_p}} \beta^\dag_{j_n}]^J_M
  \\
  +
   \phi^{J i}_{j_pj_n}[\beta_{\overline{\jmath_p}}\beta_{\overline{\jmath_n}}]^J_M+
   \widetilde\phi^{J i}_{j_p j_n}[\widetilde\beta_{j_p}\widetilde\beta_{j_n}]^J_M+
   i\xi^{J i}_{j_p j_n}[\beta_{\overline{\jmath_p}}\widetilde\beta_{j_n}]^J_M+
   i\widetilde\xi^{J i}_{j_pj_n}[\widetilde\beta_{j_p} \beta_{\overline{\jmath_n}}]^J_M
   \Bigr),
\end{multline}
The physical meaning of different terms in this definition is
explained in~\cite{Dzhioev_PRC81, Dzhioev_PRC92}. Here we just
mention that due to negative-energy tilde thermal quasiparticles
in~\eqref{ch_phonon}, the spectrum of thermal charge-exchange
phonons contains negative-energy and low-energy states which do not
exist at zero temperature. These ``new'' phonon states are
interpreted as thermally unblocked transitions between nuclear
excited states.

To find the energy and the structure of the thermal phonons we apply the equation of motion method
\begin{equation}\label{eq_mot}
\langle|\delta Q,[{\cal H},Q^\dag]]|\rangle = \omega(T)\langle|[\delta Q,Q^\dag]|\rangle
\end{equation}
under two additional constraints: (a) the phonon operators obey Bose
commutation  relations, and (b) the phonon vacuum obeys the thermal
state condition~\eqref{TSC}. The first constraint is equivalent to
averaging with respect to the BCS thermal vacuum in the equation of
motion and it leads to an orthonormality condition for the phonon
amplitudes
 \begin{multline}\label{constraint}
 \sum_{j_p j_n}\Bigl(
  \psi^{J i}_{j_pj_n}\psi^{J i'}_{j_pj_n}+
 \widetilde\psi^{J i}_{j_pj_n}\widetilde\psi^{J i'}_{j_pj_n}+
  \eta^{J i}_{j_pj_n}\eta^{J i'}_{j_pj_n}+
  \widetilde\eta^{J i}_{j_pj_n}\widetilde\eta^{J i'}_{j_pj_n}\\
  -\phi^{J i}_{j_pj_n}\phi^{J i'}_{j_pj_n}-
  \widetilde\phi^{J i}_{j_pj_n}\widetilde\phi^{J i'}_{j_pj_p}-
   \xi^{J i}_{j_pj_n} \xi^{J i'}_{j_pj_n}-
  \widetilde\xi^{J i}_{j_pj_n}\widetilde\xi^{Ji'}_{j_pj_n}\Bigr)=
      \delta_{ii'}.
      \end{multline}
The last assumption yields the following relations between
amplitudes:
\begin{align}\label{constraint1}
  \binom{\widetilde\psi}{\widetilde\phi}^{J i}_{j_pj_n}&=\frac{y_{j_p}y_{j_n}-\mathrm{e}^{-\omega_{Ji}/2T}x_{j_p}x_{j_n}}{\mathrm{e}^{-\omega_{Ji}/2T}y_{j_p}y_{j_n}-x_{j_p}x_{j_n}}
\binom{\phi}{\psi}^{J i}_{j_pj_n},
\notag \\
 \binom{\widetilde\eta}{\widetilde\xi}^{J i}_{j_pj_n}&=\frac{y_{j_p}x_{j_n}-\mathrm{e}^{-\omega_{Ji}/2T}x_{j_p}y_{j_n}}{\mathrm{e}^{-\omega_{Ji}/2T}y_{j_p}x_{j_n}-x_{j_p}y_{j_n}}
\binom{\xi}{\eta}^{J i}_{j_pj_n}.
\end{align}
Here, $x_j$ and $y_j$ ($x^2_j+y^2_j=1$)  are the coefficients of the
so-called thermal transformation which establishes a connection
between Bogoliubov and thermal quasiparticles. Note that $y_j$ are
given by the nucleon Fermi-Dirac function and they define a number
of thermally excited Bogoliubov quasiparticles in the thermal
vacuum~(see~\cite{Dzhioev_PRC81} for more details).

To derive the TQRPA equations it is convenient to introduce the
following linear combinations of amplitudes:
 \begin{align}\label{gw}
 \binom{g}{w}^{J i}_{j_pj_n}&=
 \psi^{Ji}_{j_pj_n}\pm\phi^{Ji}_{j_pj_m},
 & \binom{\widetilde g}{\widetilde w}^{J i}_{j_pj_n}&=
 \widetilde\psi^{J i}_{j_pj_n}\pm\widetilde\phi^{Ji}_{j_pj_n},
 \notag\\[3mm]
\binom{t}{s}^{Ji}_{j_pj_n}&=
 \eta^{J i}_{j_pj_n}\pm\xi^{J i}_{j_pj_n},
 & \binom{\widetilde t}{\widetilde s}^{J i}_{j_pj_n}&=
 \widetilde\eta^{J i}_{j_pj_n}\pm\widetilde \xi^{Ji}_{j_pj_n}.
\end{align}
Then from~\eqref{constraint1} it follows that
\begin{align}\label{gwgw}
  \binom{g}{w}^{J i}_{j_pj_n}&= (x_{j_p}x_{j_n}-\mathrm{e}^{-\omega_{Ji}/2T}y_{j_p}y_{j_n})\binom{G}{W}^{J i}_{j_pj_n}
  \notag\\
  \binom{\widetilde g}{\widetilde w}^{J i}_{j_pj_n}&= \mp(y_{j_p}y_{j_n}-\mathrm{e}^{-\omega_{Ji}/2T}x_{j_p}x_{j_n})\binom{G}{W}^{Ji}_{j_pj_n}
  \notag\\
  \binom{t}{s}^{Ji}_{j_pj_n}&=(x_{j_p}y_{j_n}-\mathrm{e}^{-\omega_{Ji}/2T}y_{j_p}x_{j_n})\binom{T}{S}^{J i}_{j_pj_n}
  \notag\\
  \binom{\widetilde t}{\widetilde s}^{Ji}_{j_pj_n}&=\mp(y_{j_p}x_{j_n}-\mathrm{e}^{-\omega_{Ji}/2T}x_{j_p}y_{j_n})\binom{T}{S}^{J i}_{j_pj_n},
\end{align}
where $G,~W,~T$ and $S$ are normalized according to
\begin{align}\label{norm}
  \sum_{j_nj_p}&
   \Bigl(G^{Ji}_{j_pj_n}W^{Ji'}_{j_pj_n}(1-y^2_{j_p}-y^2_{j_n})
  \notag\\
  &- T^{Ji}_{j_pj_n}S^{Ji'}_{j_pj_n}(y^2_{j_p}-y^2_{j_n})\Bigr)
  =\delta_{ii'}/(1-\mathrm{e}^{-\omega_{Ji}/T}).
\end{align}

From the equation of motion~\eqref{eq_mot} we  get the system of
TQRPA equations for unknown variables $G,~W,~T$, and $S$ and phonon
energies
\begin{align}\label{eqTQRPA}
 &G^{Ji}_{j_pj_n}\pm W^{Ji}_{j_pj_n} =\frac{2\hat J^{-2}}{\varepsilon^{(+)}_{j_pj_n}\mp\omega_{Ji}}\sum_{n=1}^{2N} d^{(Jn)}_{j_pj_n} \kappa_1^{(n)}
    \bigl(u^{(+)}_{j_pj_n}D^{Jin}_{+} \pm u^{(-)}_{j_pj_n}D^{Jin}_{-}\bigr)
    \notag\\
 &T^{Ji}_{j_pj_n}\pm S^{Ji}_{j_pj_n} =\frac{2\hat J^{-2}}{\varepsilon^{(-)}_{j_pj_n}\mp\omega_{Ji}}\sum_{n=1}^{2N} d^{(Jn)}_{j_pj_n} \kappa_1^{(n)}
    \bigl(v^{(-)}_{j_pj_n}D^{Jin}_{+} \pm
    v^{(+)}_{j_pj_n}D^{Jin}_{-}\bigr),
\end{align}
where
\begin{align}
  D^{Jin}_{+} = \sum_{j_pj_n} d^{(Jn)}_{j_pj_n}\Bigl\{ u^{(+)}_{j_pj_n}(1-y^2_{j_p}-y^2_{j_n}) G^{Ji}_{j_pj_n} - v^{(-)}_{j_p j_n}(y^2_{j_p}-y^2_{j_n})T^{Ji}_{j_pj_n}\Bigr\},
  \notag\\
  D^{Jin}_{-} = \sum_{j_pj_n} d^{(Jn)}_{j_pj_n}\Bigl\{ u^{(-)}_{j_pj_n}(1-y^2_{j_p}-y^2_{j_n}) W^{Ji}_{j_pj_n} - v^{(+)}_{j_p j_n}(y^2_{j_p}-y^2_{j_n})S^{Ji}_{j_pj_n}\Bigr\}.
\end{align}
In the above equation we have introduced  the following linear
combination of the Bogoliubov $(u,v)$ coefficients:
$u^{(\pm)}_{j_pj_n}= u_{j_p}v_{j_n}\pm v_{j_p}u_{j_n}$,
 $v^{(\pm)}_{j_pj_n}= u_{j_p}u_{j_n}\pm v_{j_p}v_{j_n}$. The factors $d^{(Jn)}_{j_pj_n}$ are given by
 \begin{equation}
   d^{(Jn)}_{j_pj_n} = \left\{\begin{array}{cc}
                f^{(Jk)}_{j_pj_n}, &   \mathrm{~if~} n = k\\
                g^{(JJk)}_{j_pj_n}, & \mathrm{~if~} n = N+k
              \end{array}\right.
 \end{equation}
 for natural parity phonons ($\pi = (-1)^J$), and
 \begin{equation}
   d^{(Jn)}_{j_pj_n} = \left\{\begin{array}{cc}
                g^{(J-1Jk)}_{j_pj_n}, & \mathrm{~if~} n = k\\
                g^{(J+1Jk)}_{j_pj_n}, & \mathrm{~if~} n = N+k
              \end{array}\right.
 \end{equation}
 for unnatural parity phonons ($\pi = (-1)^{J+1}$).

Because of the separable form of the residual  interaction the TQRPA
equations can be reduced to the set of equations for
$D^{Jin}_{\mp}$
 \begin{equation}
  \left(\begin{array}{cc}
     \mathcal{M}_1-\frac12I & \mathcal{M}_2 \\
     \mathcal{M}_2 & \mathcal{M}_3-\frac12I
   \end{array}\right)\left(
   \begin{array}{c}
     D_+ \\  D_-
   \end{array}\right)=0.
\end{equation}
The matrix elements of the $2N\times2N$ matrices $\mathcal{M}_\beta$
are the following:
\begin{eqnarray*}
\mathcal{M}^{nn'}_{1,3}&=&\frac{\kappa^{(n')}_1}{\hat J^2}\!\sum_{j_p j_n}
   d^{(Jn)}_{j_p j_n}d^{(Jn')}_{j_p j_n}\left\{
      \frac{\varepsilon_{j_pj_n}^{(+)}(u^{(\pm)}_{j_pj_n})^2}
       {(\varepsilon_{j_pj_n}^{(+)})^2-\omega^2_{Ji}}(1\!-\!y^2_{j_p}\!-\!y^2_{j_n})-
      \frac{\varepsilon_{j_pj_n}^{(-)}(v^{(\mp)}_{j_pj_n})^2}
       {(\varepsilon_{j_pj_n}^{(-)})^2-\omega^2_{Ji}}(y^2_{j_p}\!-\!y^2_{j_n})\right\},\\
\mathcal{M}^{nn'}_{2}&=&\frac{\kappa^{(n')}_1}{\hat J^2}\,\omega_{Ji}\sum_{j_n j_p}
  d^{(Jn)}_{j_p j_n}d^{(Jn')}_{j_p j_n}\left\{
  \frac{u^{(+)}_{j_pj_n}u^{(-)}_{j_pj_n}}
       {(\varepsilon_{j_pj_n}^{(+)})^2-\omega^2_{Ji}}(1-y^2_{j_p}-y^2_{j_n})-
  \frac{v^{(+)}_{j_pj_n}v^{(-)}_{j_pj_n}}
       {(\varepsilon_{j_pj_n}^{(-)})^2-\omega^2_{Ji}}(y^2_{j_p}-y^2_{j_n})\right\},
\end{eqnarray*}
where $1\le n,n'\le2N$. Thus, the TQRPA eigenvalues $\omega_{Ji}$ are the roots of the secular equation
   \begin{equation}
 \mathrm{det} \left(\begin{array}{cc}
     \mathcal{M}_1-\frac12I & \mathcal{M}_2 \\
     \mathcal{M}_2 & \mathcal{M}_3-\frac12I
   \end{array}\right)=0,
\end{equation}
while  the phonon amplitudes corresponding the TQRPA eigenvalue
$\omega_{Ji}$ are determined by Eqs.~\eqref{gw}, \eqref{gwgw}, and
\eqref{eqTQRPA}, taking into account the normalization
condition~\eqref{norm}.


\clearpage

\clearpage

\clearpage

\clearpage

\end{document}